\documentclass{ws-mpla}                 

\newcommand{\qq}{{\overline q}q}
\newcommand{\uu}{${\overline u}u \ $}
\newcommand{\dd}{${\overline d}d \ $}

\newcommand{\KK}{{\overline K}K}
\newcommand{\st}{${\overline s}s \ $}
\newcommand{\be}{\begin{equation}}
\newcommand{\ee}{\end{equation}}
\newcommand{\bea}{\begin{eqnarray}}
\newcommand{\eea}{\end{eqnarray}}

\usepackage{txfonts}                   

\begin{document}
\markboth{M.R.~Pennington}{Two photon coupling of the sigma}

\catchline{}{}{}{}{}


\title{Location, correlation, radiation:\\ where is the $\sigma$, what is its structure and what is its coupling to photons?}

\author{M.R. Pennington}

\address{Institute for Particle Physics Phenomenology, Durham University, Durham DH1 3LE, U.K. m.r.pennington@durham.ac.uk}

\maketitle
\vspace{-5.mm}
\begin{center}
Received: DAY MONTH YEAR
\end{center}

\begin{abstract}
Scalar mesons are a key expression of the infrared regime
of QCD. The lightest of these is the $\sigma$. Now that its pole in the complex energy plane has been precisely located, we can ask whether this state 
is transiently ${\overline q}q$ or 
${\overline {qq}} qq$ or  a multi-meson molecule or largely glue? The 
 two photon decay of the $\sigma$ can, in principle, discriminate between these possibilities. We review here how the $\gamma\gamma\to\pi^+\pi^-$, $\pi^0\pi^0$ cross-sections can be accurately computed. The result not only agrees with experiment, but definitively fixes the radiative coupling of the $\sigma$. This equates to a two photon width  of $(4.1 \pm 0.3)$ keV, which accords with the simple non-relativistic quark model expectation for a \uu, \dd scalar. Nevertheless, robust predictions from relativistic strong coupling QCD are required for each of the possible compositions before we can be sure which one really delivers the determined $\gamma\gamma$ coupling.
\end{abstract}



\section{Location: the vacuum of QCD and the spectrum of scalars}

\noindent 
The nature and role of scalar mesons is one of the most intriguing enigmas of strong dynamics~\cite{klempt}.
 The scalars are special: they constitute the Higgs sector of the strong interaction, giving masses to all of the light hadrons. They are in turn intimately tied to the structure of the QCD vacuum. To understand this we have to go back 60 years.
We know that just counting quarks is sufficient to explain why a typical meson, like the $\rho$, and the lightest baryon, the nucleon, have masses in the ratio of $\,2:3$. Then the question is why are pions so very light? Indeed, since  the current masses of the {\it up} and {\it down} quarks are just a few MeV, one could equally ask why are the masses of typical mesons and baryons so heavy!
The answer is in the structure of the QCD vacuum. The vacuum of QED is essentially empty with just perturbatively calculable corrections from virtual photon emission and absorption, including $e^+e^-$ pairs. The vacuum  of QCD is quite different. While asymptotic freedom ensures that quarks propagate freely over very short distances deep inside a hadron, little perturbed by the sea of quark-antiquark pairs and the clouds of gluons, over larger distances (distances of the order of a fermi), the interactions between quarks and gluons become so strong that they produce correlations in the vacuum. Correlations that we know as condensates. It is by moving through this complex vacuum that the light quarks gain their mass.

We are, of course, familiar with the idea that the world of hadron physics reflects the symmetry properties of the underlying world of quarks and gluons.
The hadron world has isospin symmetry because of the near equal mass of the {\it up} and {\it down} quarks and their flavour blind interactions. The fact that the current masses of these quarks are so much lighter than the natural scale of QCD, {\it viz.} $\Lambda_{QCD}$, means that we can regard them as massless. Then their helicity becomes a well-defined quantum number and the left and right-handed worlds decouple. QCD has an almost exact $SU(2) \times SU(2)$ chiral symmetry. In the world of hadrons this symmetry is not apparent. Scalars and pseudoscalars, vectors and axial-vectors are not degenerate in mass with closely related interactions. Consequently this symmetry must be broken. Long long ago before QCD was uncovered, Nambu proposed a mechanism for this spontaneous breakdown of chiral symmetry~\cite{nambu,njl,gml}. If the potential generated by the interactions of a scalar ($\sigma$) field and the pseudoscalar ($\pi$) has a Mexican hat shape, then nature chooses a ground state in which pions correspond to the quantum fluctuations round the bottom of the hat and are massless. In contrast, the physical scalar field, produced  by fluctuations about its non-zero vacuum expectation value are up and down the sides of the hat, implying it is massive. 
What is happening in the underlying world of QCD is that the operators $\qq$, ${\overline q} G q$, {\it etc.}, gain non-zero vacuum expectation values. To realise 
Nambu's picture, this dynamical breaking must be dominated by the $\qq$ condensate.

This is all very well-known. What is relatively new, is that these ideas have been tested in experiment and found to agree exactly with this picture. For twenty years QCD sum-rules have indicated that the value of $\langle\ \qq \, \rangle\,\simeq\,-(240$ MeV$)^3$ and it is this 240 MeV that sets the scale for the constituent mass of the {\it up} and {\it down} quarks.
Similarly, Schwinger-Dyson calculations of the Euclidean momentum dependence of the quark mass function show that such a value for the chiral condensate follows just from $\Lambda_{QCD}$. Because pions are the Goldstone bosons of chiral symmtery breaking, their interactions at low energy directly reflect the value of this condensate. Such low energy interactions can be studied experimentally in $K_{e4}$  
decay for instance. There a $K$ decays into an electron and its neutrino, plus two pions, which once formed interact strongly. Because pions are the lightest of all hadrons, their interactions are universal. Thus from the $K_{e4}$ decay distribution we can learn about the phase of the relevant $\pi\pi$ interaction.
The decay depends on five kinematic variables~\cite{cabibbo}: the di-lepton mass, the di-pion mass, the three angles, $\theta_{\ell}$, $\theta_{\pi}$ and $\phi$. In the kaon rest-frame the direction of the back-to-back di-lepton and di-pion systems defines a direction. Then in the di-lepton rest frame, each lepton goes off at an angle, $\theta_{\ell}$, to this direction, and the pions in the di-pion rest frame at an angle $\theta_{\pi}$. The plane of the two leptons and the plane of the two pions are at angle $\phi$ to each other. The 5-dimensional decay distribution then determines the relative phase of the $\pi\pi$ $S$ and $P$-wave interactions as a function of di-pion mass.  The first high statistics measurement came 6 or 7 years ago from E865 at BNL~\cite{e865}. These showed that the low energy behaviour of this phase difference corresponded to a $\qq$ condensate of exactly the size expected~\cite{colangelo}. This dominance of the $\qq$ condensate has been further checked by the preliminary results from NA48~\cite{na48}.

\begin{figure}[t]
\begin{center}
~\epsfig{file=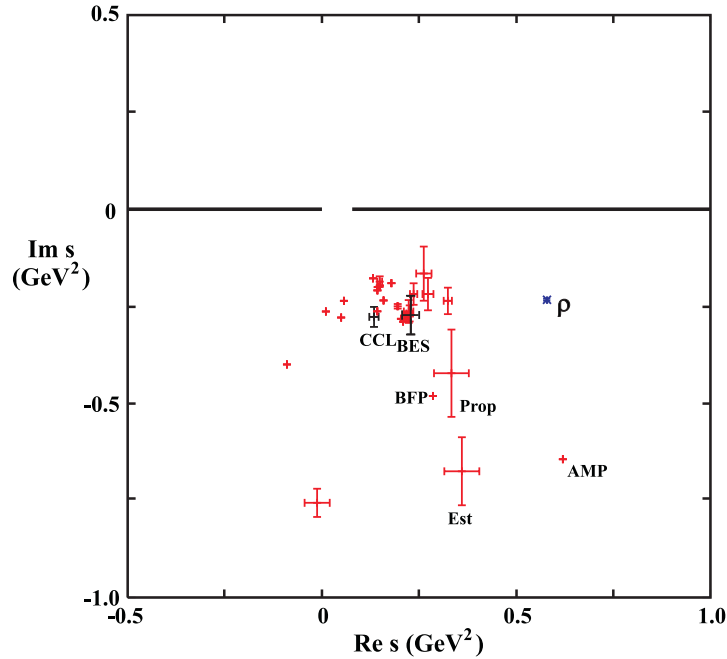,angle=0,width=9.5cm}
\caption{The complex energy squared ($s$) plane is displayed. 
The position of the $T$-matrix pole for the $\sigma$ determined in
 each of the analyses listed in the PDG tables~\protect\cite{pdg} is shown. 
The position of the $\rho$-pole is marked for orientation.
Singled out are those  labelled Prot~\protect\cite{protopopescu}, BFP~\protect\cite{bfp}, 
Est~\protect\cite{estabrooks} and AMP~\protect\cite{amp} from 30 years ago. CCL from the more 
recent Roy equation analysis~\protect\cite{ccl} and  BES from a fit to the BES 
$J/\psi \to\omega(\pi\pi)$ data ~\protect\cite{bes} are highlighted. The PDG 
listings~\protect\cite{pdg} should be seen for the complete list of other references.} 
\end{center}
\vspace{-2.mm}
\end{figure}

Thus we see that Nambu's picture of the breaking of chiral symmetry is indeed realised in nature. So  what is the scalar field,
the non-zero vacuum expectation value of which generates masses for all light hadrons?  Do we identify Nambu's $\sigma$ with the lightest $I=J=0$ state, the $f_0(600)$~? That such a meson might exist was proposed more than 40 years ago. However, 
Breit-Wigner fits to $\pi\pi$ mass distributions have found a wide range of masses and widths over the years, making it far from certain that such a state really existed. For instance, the AMP analysis~\cite{amp}, the most complete in the mid '80's, found a pole at $E= 0.87 -i 0.37$ GeV. A later treatment by Anisovich and Sarantsev~\cite{as} which included fits to the  production of two pion subsystems in ${\overline p}p$ annihilation from Crystal Barrel, 
found no pole whatsover. In contrast, more recently simple Breit-Wigner fits to $\pi^+\pi^-$ final states in $D^+\to\pi^+\pi^-\pi^+$  data from E791~\cite{e791}  and in $J/\psi\to \omega \pi^+\pi^-$ from BESII~\cite{bes} have found poles at 
$E= (478\pm 29) - i(162 \pm 11)$ MeV and $E= (541 \pm 29) - i(252 \pm 42)$ MeV, respectively.  However, these fits for the most part ignore the related phase variation determined  for $\pi\pi\to\pi\pi$ in the elastic region. The scatter of pole positions from all the analyses quoted in the latest PDG tables~\cite{pdg,protopopescu,bfp,estabrooks,amp,ccl,bes} are shown in the complex $s = E^2$ plane in Fig.~1.
\begin{figure}[t]
\begin{center}
~\epsfig{file=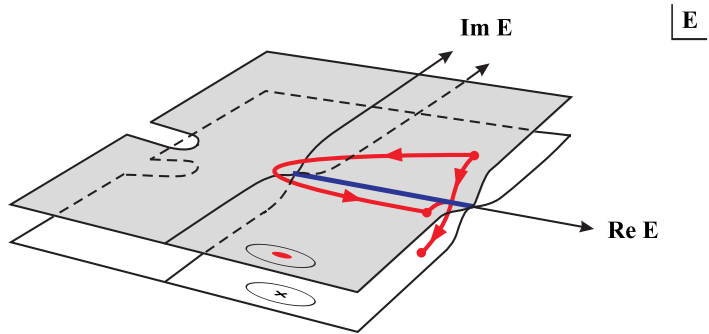,width=9.3cm}
\vspace{2mm}
\caption
{An illustration of the sheets and cut structure of the complex energy plane
in a world with just one threshold and how these are connected. 
This represents the structure relevant to $\pi\pi$ scattering near its threshold. Experiment is performed on the top shaded sheet, just above the cut along the real energy axis. The {\it cross} on the lower, or second, sheet indicates where the $\sigma$-pole resides at $E^2 = s =s_R$. The {\it ellipse} above this on the top, or first, sheet indicates where the $S$-matrix is zero. } 
\end{center}
\vspace{-2.mm}
\end{figure}

Now we know, thanks to a rather precise analytic continuation into the complex energy plane by Caprini, Colangelo and Leutwyler~\cite{ccl}, that a pole does exist, signalling a state in the spectrum of hadrons. The analytic continuation starts with fixed-$t$ dispersion relations. As always the left hand cut requires knowledge of crossed-channel exchanges. The unique feature of $\pi\pi$ scattering is its 3-channel crossing symmetry. This means both the right and left hand cuts involve integrals over $\pi\pi\to\pi\pi$ scattering.
Making use of this crossing property, and then partial wave projecting, leads to the Roy equations~\cite{roy}. Solving this set of rigorous equations inputting experimental information above 800 MeV, together with chiral constraints, allows the $I=J=0$ $\pi\pi$ partial wave to be determined everywhere on the first sheet of the energy plane~\cite{cgl}, Fig.~2.
As shown by Caprini {\it et al.}~\cite{ccl}, this  fixes a zero  of the $S$-matrix (symbolically depicted by the solid ellipse in Fig.~2) at $E\,=\,441\,-i\,227$ MeV, which reflects a pole (denoted by the cross) on the second sheet at the same position. This not only confirms the $\sigma$ as a state in the spectrum of hadrons but locates the position of its pole very precisely with errors of only tens of MeV. This is within the region found by Zhou {\it et al}~\cite{zhou}, who also took into account crossing and the left-hand cut. While the chiral expansion of amplitudes can have no poles at any finite order, particular summations may. The inverse amplitude method~\cite{whs99} is one such procedure. Application of this  by Pelaez {\it et al.}~\cite{pelaez1} did indeed find a pole in the same domain as Caprini {\it et al.} but many years earlier. However, without a proof that the Inverse Amplitude Method, rather than any other, provided precision unitarisation of the low order chiral expansion, the present author rather believed the analysis of a wide range of data 
of Ref.~10 that indicated no pole (or perhaps a very distant one). Now we know differently. There is a pole with a well-defined location. 
This is far from the position proposed by the treatment of Ishida {\it et al.}~\cite{ishida1,ishida2,ishida3,ishida4,ishida5}, the deficiencies of which were explained long ago in Ref.~21.
 (Bugg~\cite{buggtak} has added to these arguments in response to the discussion on the position of the $\kappa$ by the same group~\cite{takamatsu} .)

\begin{figure}[t]
\begin{center}
~\epsfig{file=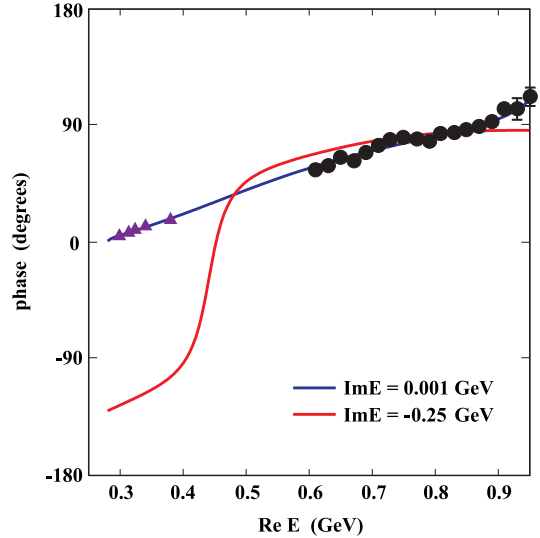,width=7.cm}
\caption
{How the phase of the $I=J=0$ $\pi\pi\to\pi\pi$ scattering amplitude varies with the real part of the energy: (i)~just above the real axis, where it is determined by experiment from the CERN-Munich group~\protect\cite{cern-munich1,cern-munich2,cern-munich3} (circles)  and $K_{e4}$ decays~\protect\cite{e865} (triangles) near threshold, (ii) at Im$E = -0.25$ GeV on the second (unphysical) sheet.}
\end{center}
\vspace{-3.mm}
\end{figure}

With a narrow resonance, there would naturally be a close correlation between
the phase variation of the underlying amplitude on the real axis and in the complex plane, as one passes the pole.
 However, for the very short-lived $\sigma$, \lq\lq deep'' in the complex energy plane, this simple connection is lost.
In Fig.~3 we show the phase of the $I=J=0$ $\pi\pi\to\pi\pi$ scattering amplitude along two lines in the complex energy plane. The phase along the real axis is compared with relevant data from scattering~\cite{cern-munich3} and $K_{e4}$ decays~\cite{e865} in Fig.~3. One sees how different this phase is compared with that on the lower sheet of Fig.~2 at Im$E=-0.25$ GeV. That deep in the complex plane shows the 180$^o$ phase change expected of a resonance. It is the dramatic variation in the amplitude as one moves away from the real axis that has made the $\sigma$'s existence and location so uncertain for so long. With the right tools for accurate analytic continuation, this is no longer an issue.

  But is this $\sigma$ the chiral partner of the $\pi$~\cite{scadron}? Is it the Higgs of the strong interaction? There exist many other isoscalar scalars $f_0(980), f_0(1370), f_0(1510), f_0(1720)$. What role do these play? What is more, these have isodoublet and isotriplet partners, like the $K_0^*(800)$ (or $\kappa$), $K_0^*(1430)$, $a_0(980)$ and $a_0(1430)$. How are these all related? The quark model would lead us to expect a $\,^3P_0\,$ nonet, but there are 19 scalars just listed: enough for two nonets with an isosinglet left over! Of course, there is a danger of subjectivity in what is included amongst this list. That the nine lightest do not fit a simple $\qq$ pattern anticipated from vector and tensor mesons is readily seen by noting that
both the $f_0(980)$ and $a_0(980)$ sit very close to the $\KK$ threshold and\ couple very strongly to these channels. In a simple quark multiplet only the \st state does that.

The possibility of discovery of pentaquark states has revived interest in colour $\overline 3$ scalar diquarks, formed from quarks of different flavour, like $[ud]$, $[us]$ and $[ds]$~\cite{jaffe,jaffe-wilczek}. Such diquarks can bind with the corresponding anti-diquarks to form tetraquark systems. The lightest multiplet is a nonet and lighter~\cite{jaffe} than the corresponding $^3P_0$ $\qq$ nonet, as in Fig.~4. Moreover, the heaviest tetraquarks  are an $f_0$ and $a_0$: ${\overline {[ns]}}[ns]$, with $n=u,d$, suggesting these might be good assignments for the $980$ MeV states. However, Weinstein and Isgur~\cite{wi1,wi2,wi3} found within their potential model that the only scalar tetraquark systems to bind would be in the form of  $\KK$-molecules. This would predict just four states, the $f_0$ and $a_0$, and not a complete nonet.   The distiction between ${\overline q}q$ and ${\overline{qq}}qq$ states is most marked in the theorists' favourite world of large $N_c$ (where $N_c$ is the number of colours). Then the $\qq$ nonet in Fig.~4 becomes stable, while the tetraquark mesons 
merge into the continuum. Of course, we do not live in a world in which $N_c$ is large. Indeed, nowhere is this more apparent than in the scalar sector.
Pelaez~\cite{pelaez2a,pelaez2b} has tracked the $\sigma$ and $\kappa$ as $N_c$ is varied and concluded that they are largely tetraquarks.
 With the addition of a glueball with mass computed on the lattice in a world without quarks to be in the region of 1.5-1.7 GeV~\cite{latticeglueball1,latticeglueball2,latticeglueball3}, we have our 19 \lq\lq lightest'' scalars. Of course, these states $\qq$, ${\overline {qq}}qq$, (or even molecules) and glueball are not orthogonal to each other. They inevitably mix.
So how can we try to distinguish which scalars have what composition?
In particular, what is the composition of the lightest scalar~\cite{bugg}, the $\sigma$?
\begin{figure}[t]
\begin{center}
~\epsfig{file=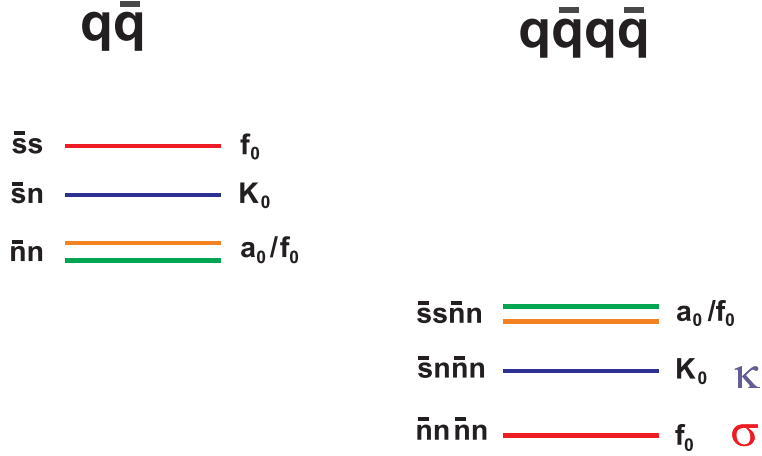,width=10.cm}
\vspace{2mm}
\caption
{The spectrum of scalar states given by a simple {\it ideal} nonet of ${\overline {q}}q$ and of ${\overline{qq}qq}$ mesons, where $n = u,d$, indicating which would be identified with the isoscalar $\sigma$ and its isodoublet partner, the $\kappa$, if the observed hadrons had these {\it orthogonal} compositions.} 
\end{center}
\end{figure}

\section{Correlation: photons probing hadron composition}

Two photon interactions can illuminate these issues.
For orientation, let us presume we have extracted the two photon couplings of resonant states. Before looking at the enigmatic scalars, let us consider 
 the two photon couplings of the tensor mesons, $f_2$, $a_2$ and $f_2'$, that belong to an ideally mixed ${\overline q}q$ multiplet. Their radiative widths are proportional to the square of the average charge squared of their constituents, Fig.~5.
The absolute scale of their couplings depends on dynamics: on how the ${\overline q}q$ pair form the hadron. In the non-relativistic limit, as with charmonia, this is simply related to the wavefunction at the origin. If we assume that these are equal for the tensor states, then we have the prediction that
\be
\Gamma(f_2\to\gamma\gamma)\;:\;\Gamma(a_2\to\gamma\gamma)\;:\;\Gamma(f_2'\to\gamma\gamma)\;=\; 25\,:\,9\,:\,2\quad .
\ee
Experiment~\cite{pdg} is in reasonable agreement with this, given the uncertainties in extracting  couplings from data (as we shall see).
If we apply the same ideas to the pseudoscalars, $\pi^0, \eta, \eta'$, though not ideally mixed, we would not be able to reproduce experiment~\cite{pdg}:
\be
\Gamma(\pi^0\to\gamma\gamma)\;:\;\Gamma(\eta\to\gamma\gamma)\;:\;\Gamma(\eta'\to\gamma\gamma)\;\simeq\; 1\,:\,60\,:\,500\quad .
\ee
\begin{figure}[t]
\begin{center}
~\epsfig{file=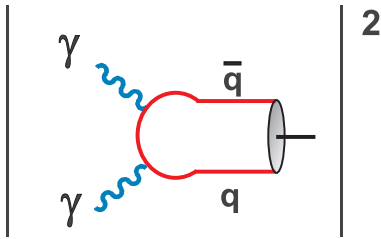,width=5.cm}
\vspace{1mm}
\caption
{Two photon decay rate of a quark model hadron is the modulus squared of the amplitude for $\,\gamma\gamma\,$ to produce a $\,{\overline q}q\,$ pair and for these to bind by strong coupling dynamics. In present calculations these are treated with crude approximations. A genuine relativistic strong coupling approach is  essential for the lightest pseudoscalars and scalars. } 
\end{center}
\vspace{-3.mm}
\end{figure}
\noindent While the non-relativistic quark model  determines the intrinsic coupling
to photons, it does not include the dynamics of binding the quark and antiquark into a meson, Fig.~5. A Lagrangian for the  pseudoscalar-photon-photon interaction would introduce a factor of mass cubed for each meson. As noted by Hayne and Isgur~\cite{hayneisgur}, such factors are just what is needed to bring the quark model prediction into agreement with experiment of Eq.~(2). However, what this  teaches us is that we should not be using the non-relativistic quark model and correcting it by factors of hundreds for light quarks. Rather we need genuinely relativistic strong coupling calculations of such radiative widths. As we shall see, this need equally applies to the scalars.

\begin{figure}[p]
\begin{center}
~\epsfig{file=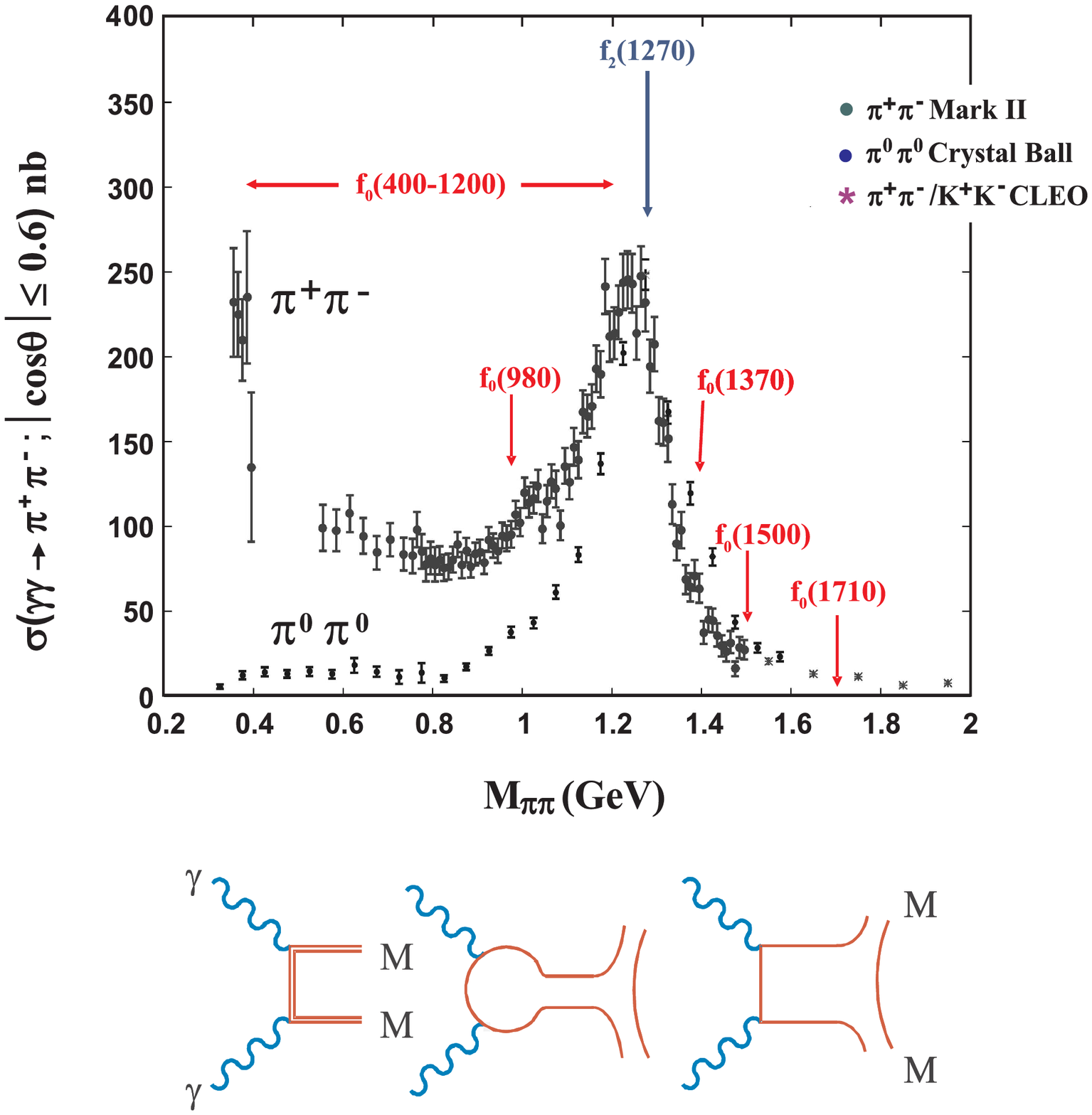,width=10.3cm}
\caption
{Integrated cross-section for $\gamma\gamma\to\pi\pi$ as a function of c.m. energy $M(\pi\pi)$ from Mark II~\protect\cite{MarkII}, Crystal Ball~\protect\cite{CB1,CB2} and CLEO~\protect\cite{cleo}. The $\pi^0\pi^0$ results have been scaled to the same angular range as the charged data and by an isospin factor. Below are graphs describing the dominant dynamics in each kinematic region,
as discussed in the text. } 
\end{center}
\vspace{1.5mm}
\begin{center}
~\epsfig{file=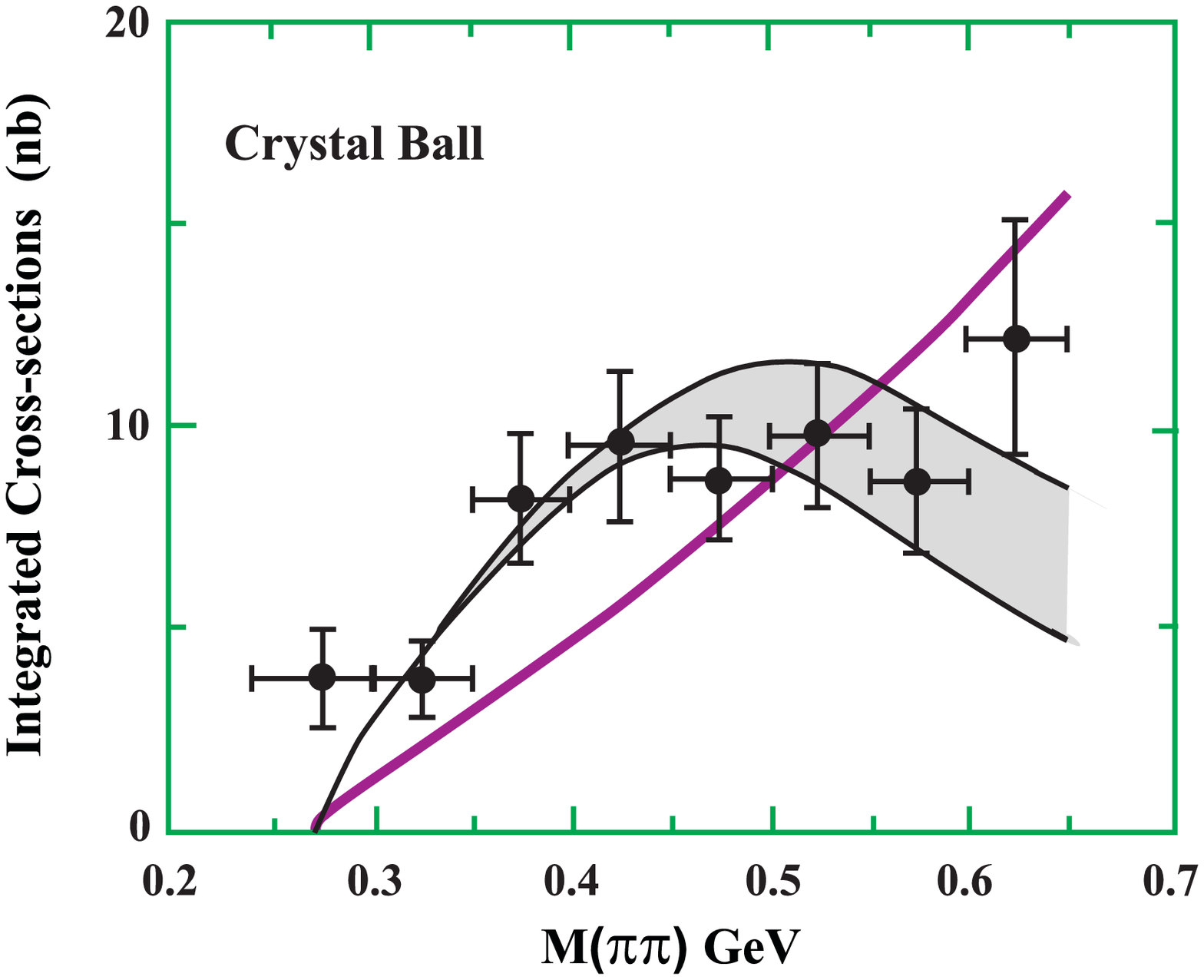,width=6.8cm}
\caption
{Cross-section for $\gamma\gamma\to\pi^0\pi^0$ integrated over $|\cos \theta^*| \le 0.8$ as a function of the $\pi\pi$ invariant mass $M(\pi\pi)$. The data are from Crystal Ball~\protect\cite{CB1,CB2}. The line is the prediction of $\chi PT$ at one loop $(1\ell)$~\protect\cite{bijnens}. The shaded band shows the dispersive prediction~\protect\cite{mp,mrpgamma1,mrpgamma2} described later --- its width reflects the uncertainties in 
experimental knowledge of both $\pi\pi$ scattering and vector exchanges.} 
\end{center} 
\vspace{-2.mm}
\end{figure}

Now how do we extract such two photon couplings?  Let us consider the exclusive process
for which we have the most information, {\it viz.} $\gamma\gamma\to\pi\pi$, with both charged and neutral pions, as displayed in Fig.~6. At low energy, the photon has long wavelength and interacts with the whole of the hadron. Consequently, it \lq\lq sees'' the charged pions, but not the neutral. $\pi^0\pi^0$ production is small. $\pi^+\pi^-$ is large (Fig.~6). How large is determined by the charge of the pion. However, as the energy increases (by just 1 GeV) the wavelength of the photon shortens sufficiently that it recognises that pions, whether charged or neutral, are made of the same charged constituents, namely quarks, and causes these to resonate. Since photons have spin-1, they readily produce the prominent tensor meson $f_2(1270)$, seen in Fig.~6. As the energy increases further, the photon probes not constituent quarks, but current quarks. Their interactions can be calculated perturbatively as shown by Brodsky and Lepage~\cite{brodsky-lepage}. Allowing for a free normalisation, then these predictions for both the energy and c.m.angular distribution agree well with the latest data from Belle~\cite{BelleMM}. 
What we are interested in here is the scalar signal underneath the large $f_2(1270)$ from $\pi\pi$ threshold upwards, Fig.~6.

To see how to extract this signal, let us focus on $\pi\pi$ production close to threshold. Since pions are the Goldstone bosons of chiral symmetry  breaking, their interactions at low energy can be computed perturbatively~\cite{GL}. For $\gamma\gamma\to\pi^0\pi^0$, the lowest order in chiral perturbation theory ($\chi PT$) involves one loop graphs which predict a cross-section rising almost linearly with energy from threshold~\cite{bijnens,donoghue}, as seen in Fig.~7. This is in rather poor agreement with the only normalised data from Crystal Ball~\cite{CB1}. This was once a little worrying, since Maiani~\cite{maianichpt} described this process as providing a \lq\lq gold-plated'' prediction of $\chi PT$, suggesting that the data might be wrong. However, one can calculate this cross-section almost exactly at low energy, by noting that
 neutral pions can be produced by first $\gamma\gamma\to\pi^+\pi^-$ by the Born amplitude and then $\pi^+\pi^-$ can interact to form the $\pi^0\pi^0$ final state~\cite{lyth,mennessier,truong,mp}. That this cross-section is calculable from strong interaction dynamics makes it almost unique among hadronic processes. This is because of Low's low energy theorem~\cite{low,mp,mrpgamma1,mrpgamma2}. This states that the amplitude for $\gamma\pi\to\gamma\pi$ at threshold is exactly given by the one pion exchange Born amplitude. Though this specifies the amplitude   at just one kinematic point, the amplitude smoothly approaches this limit throughout the low energy region. This is because of the closeness of the $t$ and $u$-channel pion poles to the physical region for all three channels. This means that the Born amplitude modified by final state interactions controls the $\gamma\gamma\to\pi\pi$ amplitude in the low energy region, where the pion pole is so much closer than any other exchange: the next nearest being the $\rho$ and $\omega$.  
\begin{figure}[b]
\begin{center}
~\epsfig{file=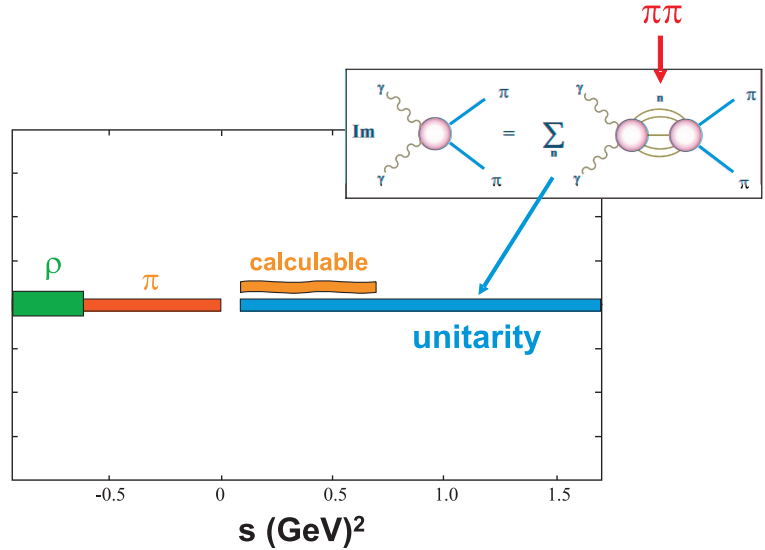,width=10.cm}
\caption{The complex $s$-plane structure of the 
$\gamma\gamma\to\pi\pi$ amplitudes. $\pi$ labels the 
 the left hand cut controlled by the pion exchange Born term,  and $\rho$
denotes where the vector exchanges contribute to the discontinuity.
 The discontinuity over the right hand cut is determined by the 
unitarity relation depicted. The $\pi\pi$ intermediate state is effectively
 the only one possible from $\pi\pi$ threhold up to the opening of the ${\overline K}K$ channel. As discussed in the text, this implies the 
$\gamma\gamma\to\pi\pi$ amplitudes is  \lq\lq exactly'' calculable in the region shown. }
\end{center}
\end{figure}
Consequently, if we consider the analytic structure of the partial wave amplitudes, Fig.~8, they have a left hand cut generated by crossed channel exchanges, in which essentially the part from $s=0$ to $s\simeq -m_{\rho}^2\,$ is known from pion exchange, and a right hand cut controlled by direct channel dynamics. (As usual $s$ is  the square of the $\gamma\gamma$ and $\pi\pi$ c.m. energy.) The discontinutity across this right hand cut is specified by unitarity:
\be
{\rm Im}\, {\cal F}(\gamma\gamma\to\pi\pi;s)\;=\;\sum_n\;{\cal F}(\gamma\gamma\to H_n;s)^*\,{\cal T}(H_n\to\pi\pi;s)\quad ,
\ee
where  ${\cal F}$, ${\cal T}$ are partial wave amplitudes with definite isospin
and the sum is over all allowed hadronic intermediate states $H_n$, see the inset in Fig.~8.  In the low energy region only $\pi\pi$ is possible. Consequently, knowing the $\pi\pi\to\pi\pi$ partial wave amplitudes 
from experiment, we can compute the $\gamma\gamma\to\pi\pi$ partial waves.

\begin{figure}[b]
\begin{center}
~\epsfig{file=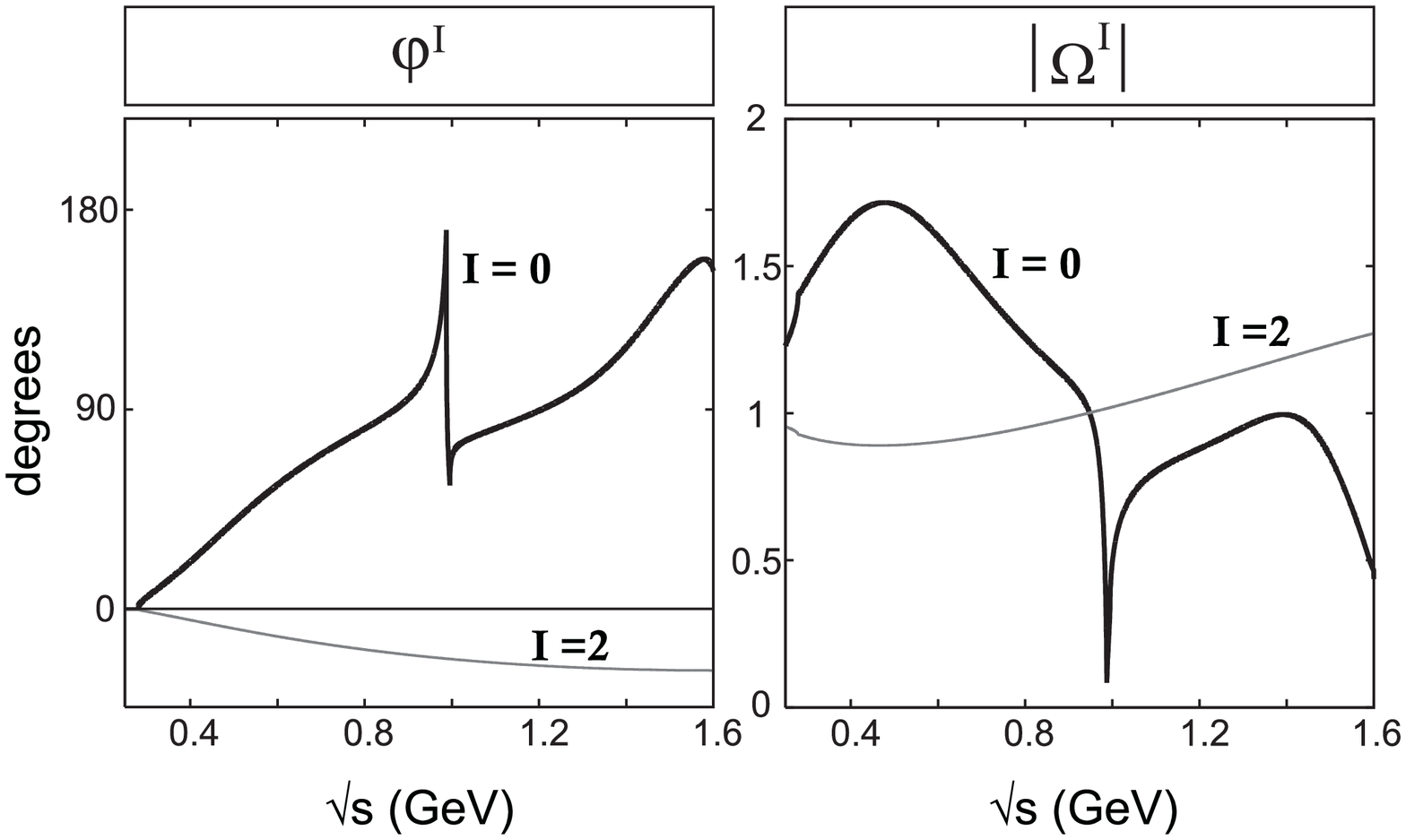,width=12.5cm}
\caption{Illustrative $I=0,\,2$ $\gamma\gamma\to\pi\pi$ $S$-wave Omn\`es functions, $\Omega ^I(s)$, phases and moduli, related by  Eq.~(4).}
\end{center}
\end{figure}
To proceed with this calculation, let
 us first consider the $S$-wave $\gamma\gamma\to\pi\pi$  amplitudes with isospin $I$, ${\cal F}^I(s)$ --- we will comment on higher partial waves later. Each of these amplitudes, with $I=0,\,2$, being complex has a phase $\phi^I(s)$ along the right hand cut, when $s$ is above the two pion threshold, {\it i.e.} $s\, >\, s_{th}\,\equiv\, 4m_{\pi}^2$.
In the elastic region, unitarity, Eq.~(3) implies Watson's theorem, requiring the phase of each of these partial waves to be
 the same
as the phase  of the corresponding $\pi\pi$ partial wave amplitude with the same spin and isospin
(in principle, up to integer multiples of $\pi$. However, at $\pi\pi$ threshold, the phases are equal to zero.)
  To implement this constraint we define the Omn\`es function, $\Omega^I(s)$,
which by construction has phase $\phi^I(s)$:
\be
\Omega^I(s)\;=\;\exp \left[ \frac{s}{\pi}\,\int_{s_{th}}^{\infty}\;
ds'\,\frac{\phi^I(s')}{s'(s'-s)} \right]\, \quad .
\ee
Thus the $\gamma\gamma\to\pi\pi$ $S$-wave amplitudes, ${\cal F}^I(s)$,
can be written as $P^I(s)\,\Omega^I(s)$, where $P^I(s)$ is a function which is real along the right hand cut with $s > s_{th}$. 
The phase, $\phi^I$, is simply the phase-shift in the region of elastic unitarity, which applies up to ${\overline K}K$ threshold, since multi-pion channels are negligible below 1.2 GeV. 
Moreover, in the low energy region of interest where $|\,s\,|\,\sim\, 0.25$ GeV$^2$, the unknown phase above 1 GeV affects the results little. This is checked by replacing the $\pi\pi\to\pi\pi$ phase with that for $\pi\pi\to {\overline K}K$. Such a change is equivalent to assuming the $\pi\pi$ final state in the two photon process is only accessed through a ${\overline K}K$ intermediate state. Outside the narrow confines of the $f_0(980)$ region, this would be an extreme possibility. Nevertheless, the effect is small and included in the uncertainties we quote.
Representative input $\pi\pi$ $S$-wave phases, $\phi^I(s)$, for $I=0,\,2$ and the resulting Omn\`es functions are shown in Fig.~9.

As already discussed the lightness of the pion means that Born amplitude not only equals
 the whole amplitude  as $s \to 0$, and $t,u \to m_{\pi}^2$, at the threshold for Compton scattering $\gamma\pi\to\gamma\pi$, but in the whole low energy region, as illustrated in Fig.~8.  To implement this, let us begin by denoting
 the left hand cut contribution to the 
$\gamma\gamma\to\pi\pi$ partial wave amplitudes generated by crossed 
channel exchanges collectively
by ${\cal L}^I(s)$. As indicated in Fig.~8,
 pion exchange determines the discontinuity in the whole region $0\,>\,s\,>\,-M_V^2$, beyond which other exchanges like $\rho, \omega$ start to contribute. While the Born term assumes pointlike couplings for the pion, any form-factor dependence only affects the left hand cut for $s\,<\, -M_V^2$, since it is vector masses, $M_V$, that set the scale for such charged radii. Consequently, the left hand cut from $s=0$ to $s\,\simeq\, -0.5$ GeV$^2$ is precisely known and that is all we require to fix the amplitude in the region of the pole at $s\,=\,s_R\,=\,E_R^2\,$ in Fig.~2. 
\begin{figure}[t]
\begin{center}
~\epsfig{file=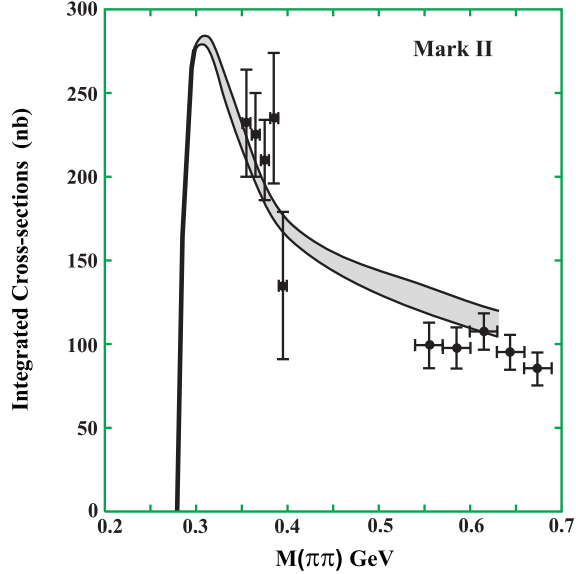,width=7.5cm}
\caption
{Cross-section for $\gamma\gamma\to\pi^+\pi^-$ integrated over
$|\cos \theta^*| \le 0.6$ as a function of the $\pi\pi$ invariant mass $M(\pi\pi)$. The data are from Mark II~\protect\cite{MarkII}. The shaded band shows the dispersive prediction~\protect\cite{mp,mrpgamma1,mrpgamma2} based on the Born term modified by calculable final state interactions in all partial waves. The width of the band reflects the uncertainties in 
experimental knowledge of both $\pi\pi$ scattering and vector exchanges.} 
\end{center}
\end{figure}

  To see how, let us construct
 the function $G^I(s)\;\equiv\;\left({\cal F}^I(s)\,-\,{\cal L}^I(s)\right)\,\Omega^I(s)^{-1}$, which only has a right hand cut. Its
discontinuity is ${\cal L}^I(s)\,\sin \phi^I(s)/| \Omega^I(s) |$, which is accurately known at low energies.  This information is embodied in a dispersion relation for the
function $G^I(s)$ using a contour round its right hand cut and closed at infinity.
 While the behaviour of $G^I(s)$ means the  integral at infinity converges with just one subtraction, it is more convenient  for our purpose to ensure 
that the integrals are dominated
by the known low energy regime of $|\, s'\, |\, <\, M_{\rho}^2$. This is achieved by making two subtractions:
\be
{\cal F}^I(s)\;=\;{\cal L}^I(s)\,+\,c_I s\, \Omega^I(s)\, 
+\,\frac{s^2}{\pi}\,\Omega^I(s)\,\int_{s_{th}}^{\infty}\, ds'\,\frac{{\cal L}^I(s')\,\sin \phi^I(s')}{s'^2\,(s' - s)\,|\,\Omega^I(s')\,|}\quad .
\ee
The constants $c_I$ are specified by the QED low energy theorem and chiral dynamics.
These two conditions apply to the amplitudes with the pions of specific charge (which are combinations of those with definite isospin).
Low's theorem, as shown by Goldberger {\it et al.}~\cite{goldberger,henry} requires that the charged pion $S$-wave amplitude ( ${\cal F}^c$ for $\gamma\gamma\to\pi^+\pi^-$), 
\be
{\cal F}^{c}(s)\;\to\; B(s)\, +\,{\cal O}(s^2)\quad {\rm as}\quad  s \to 0 ,
\ee 
where $B(s)$ is the Born $S$-wave, while chiral dynamics requires that the neutral pion $S$-wave amplitude (${\cal F}^n$ for $\gamma\gamma\to\pi^0\pi^0$)
\be
{\cal F}^{n}(s)\;=\;0\quad
 {\rm at}\quad s\,=\,{\cal O}(m_{\pi}^2)\quad .
\label{sn}
\ee
At one loop level in Chiral Perturbation Theory~\cite{donoghue},
$
{\cal F}^{n}(s)\;\propto \,{\cal T}(\pi^+\pi^-\to\pi^0\pi^0)
$
and so places the Adler zero exactly at $s\,=\,m_{\pi}^2$ at this order. However, its precise position hardly affects our results. 
An analogous procedure fixes the higher waves too. Their threshold behaviour is enough to fix their subtraction constants~\cite{mp,mrpgamma1,mrpgamma2}. Of these waves, the $D$-wave with helicity two is critical to the calculation of observables~\cite{mrpgamma1,mp}.

In this way the  $\gamma\gamma\to\pi\pi$ cross-section in the low energy region
is absolutely fixed. Precision comes from the accurate determination of the $\pi\pi$ $S$-wave amplitudes~\cite{boglione1,boglione2} obtained by combining new results from decays like $K_{e4}$, $J/\psi \to \phi X$ and $D\to\pi X$ ($X = \pi\pi, {\overline K}K$) with the Roy equations, which fixes the pole at $s=s_R$. 
This calculation reproduces the cross-section for the production of charged and neutral pions as measured by Mark~II~\cite{MarkII} and Crystal Ball~\cite{CB1}, respectively, in the low energy region with no free parameters, as displayed in Figs.~7,~10.
 The bands shown delineate the uncertainties due to (i)~different $\gamma\gamma$ phases $\phi^I(s)$ above ${\overline K}K$ threshold and (ii)~different positions of the Adler zero in Eq.~(7).  Notice that the cross-section is very  nearly  unique up to 450 MeV.
That the $\gamma\gamma\to\pi\pi$ amplitudes are reliably calculable at low energy from the Born amplitude modified by experimentally determined final state interactions has two important applications.

\section {Radiation: the two photon coupling of the $\sigma$}

Of course, the $I=0$ $\pi\pi$ phase and Omn\`es function, shown in Fig.~9, 
know about the $\sigma$-pole at $s=s_R$
deep in the complex plane close to both the right and left hand cuts of Figs.~2 and 8.
  Not only can we determine the $\gamma\gamma$ amplitudes ${\cal F}^I(s)$ along the upper side of the right hand cut on the physical sheet where experiments are performed, but everywhere on this first sheet shown shaded in Fig.~2. In particular, we can determine the $I=0$ amplitude 
at $s\,=\,s_R$, marked by the dot and cross in Fig.~2.

The right hand cut structure, Fig.~8, of the $\gamma\gamma\to\pi\pi$ amplitude mirrors that of the corresponding hadronic amplitude, ${\cal T}^I$, for  $\pi\pi\to\pi\pi$ in the region of elastic unitarity, so that:
\be
{\cal F}^I(s)\;=\;\alpha^I(s)\;{\cal T}^I(s)\quad .
\ee
This satisfies Eq.~(3) if $\alpha^I(s)$ is a real function for $s > 0$. The function, $\alpha^I(s)$, represents
the intrinsic coupling of $\gamma\gamma\to\pi\pi$, while the $\pi\pi\to\pi\pi$ amplitude, ${\cal T}$, describes the final state interactions, which colour and shape the electromagnetic process~\cite{amp}. The function $\alpha^I(s)$ is determined by crossed-channel dynamics  --- the $N$ of $N/D$ for the {\it cogniscienti}.
 
   At $s\,=\,s_R\,$ on the first sheet, the amplitude ${\cal T}^{I=0}(s)\,=\,i/2\rho(s)$, since the $S$-matrix element vanishes at this point. $\rho(s)$ is, as usual, the phase-space factor
$\rho(s)\,=\,\sqrt{1 - s_{th}/s}$. The dispersion relation on the first sheet then determines
the coupling function $\alpha(s_R)$, which not having a right hand cut, has the same value on the second sheet of Fig.~2. 

Let us introduce subscripts to label the sheets $I$ and $II$, while the superscripts continue to denote isospin.
 In the neighbourhood of the pole on the second sheet, the $\gamma\gamma\to\pi\pi$ $S$-wave amplitude is given by
\be
{\cal F}^0_{II}(s) \;\simeq\;\frac{g_{\gamma}\,g_{\pi}}{s_R - s}\; ,\quad{\rm while}\quad
{\cal T}^0_{II}(s) \;\simeq\;\frac{g_{\pi}^2}{s_R - s}\; .
\ee
Even if factorised residues are not strictly appropriate for such a very short-lived state, Eqs.~(10,11) below  provide a physically meaningful and unambiguously defined~\cite{whs99} $\gamma\gamma$ width.
$\alpha^0(s_R)$ determines the ratio of $g_{\gamma}/g_{\pi}$ for the isoscalar resonance. 
Now the hadronic amplitude on sheet I is related to that on sheet II by
\be
\frac{1}{{\cal T}_{II}(s)}\;=\;\frac{1}{{\cal T}_{I}(s)}\,+\,2\,i \rho\quad ,
\ee
so that
\be
g_{\gamma}^2\;=\;\lim_{s\to s_R}\;\frac{(s - s_R)\,F^0_{I}(s)^2}{\left({\cal T}^0_{I}(s)\,-\,i/2\rho\right)}\quad .
\ee
The $I=0$ amplitude ${\cal T}^0$ encodes the
key $\pi\pi$ final state interactions and so automatically incorporates the effect of the $\sigma$. Indeed, 
the representation, cited above~\cite{boglione1,boglione2} for ${\cal T}^0$, has a pole at $E\,=\,441\,-i\,227$ MeV, within the error ellipse found by Caprini {\it et al.}~\cite{ccl}. Inputting this hadronic amplitude on sheet I into the present dispersive calculation then determines the residue of the $\sigma$-pole in $\gamma\gamma\to\pi\pi$, and in turn specifies~\cite{morpenngg,mp2} its radiative width to be
\be
\Gamma(\sigma\to\gamma\gamma)\;=\; \frac{\alpha^2 |\,\rho(s_R)\,g_{\gamma}^2\,|}{4 M_{\sigma}}\,=\,(4.1\,\pm\,0.3)\; {\rm keV}\;,
\ee
as found in Ref.~69.

The low energy theorem and knowledge of $\pi\pi$ scattering has allowed us to determine the low energy $\gamma\gamma\to\pi\pi$ partial waves that accurately sum to describe both the charged and neutral  data of Figs.~6,~7 and~10. In these data the broad scalar is far from apparent. The near threshold peak in the charged cross-section is due to one pion exchange Born term, modified by final state interactions. The presence of this Born term, means there are large $I=0$ and $I=2$ components in the $\gamma\gamma$ amplitude. These essentially add in the charged pion channel and subtract in the neutral one. It is this that makes the \lq\lq $\sigma$'' component far from obvious in the data.
Nevertheless, one naturally asks:
  how does the large radiative width for the $\sigma$ of 4 keV  square with the size of the $\gamma\gamma\to\pi\pi$ cross-sections shown in Figs.~7,~10. Simplistic reasoning would  totally {\it ignore} the key requirement that final state interactions shape the $\pi\pi$ distribution in a well-defined way. Then one would say that, with no Born contribution in the neutral pion channel, the cross-section should directly reflect the appearance of resonant structures. If this is the $\sigma$, then one can read off, from the observed cross-section in Fig.~7 of 10 to 12 nanobarns,  a $\gamma\gamma$ width an order of magnitude smaller than we have deduced.
However, this is too naive.

The reason is that final-state interactions are crucial. These apply quite differently to the $I=0$ and 2 $\gamma\gamma$ amplitudes. In hadronic reactions $I=2$ amplitudes, being {\it exotic} in the quark model, are much smaller than those with $I=0$. In contrast in the two photon
process both $I=0,\,2$ are equally important. The $\sigma$  appears in the $I=0$ amplitude, and this can only be separated from data by analysing $\gamma\gamma\to\pi^+\pi^-$ and $\pi^0\pi^0$ together~\cite{morpenngg,mp2}.

\begin{figure}[t] 
\begin{center}
\epsfig{file=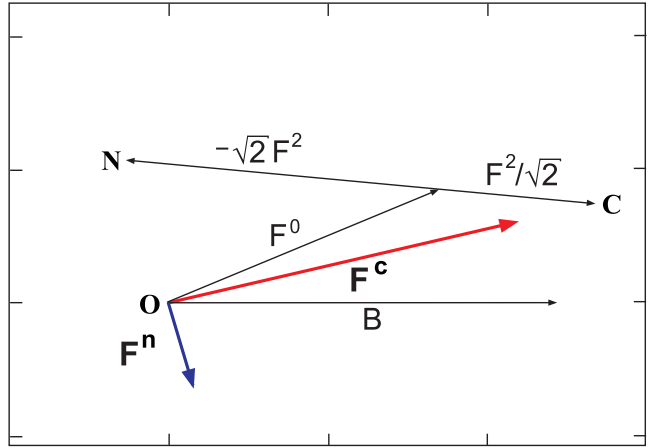,width=8.5cm}
\caption{$\gamma\gamma\to\pi\pi$ $S$-wave amplitudes, $F$, at 400 MeV with definite isospin and with definite charges as indicated by the superscripts. $B$ is the Born amplitude $S$-wave for comparison. $OC$ and $ON$ define the directions of the charged and neutral pion amplitudes as given by the vector sums described in the text.}
\end{center}
\end{figure}

As we have seen what is really happening is that the Born amplitude is modified by final state interactions to ensure Watson's theorem is satisfied. As a result the
$I=0$~and~$2$ $S$-wave amplitudes are no longer real and exactly cancelling in the neutral channel. A vectorial representation of this is shown at 400 MeV in Fig.~11.
The $I=0$ component has the phase of $I=0$ $S$-wave $\pi\pi$ scattering, while that with $I=2$ has the phase of the corresponding isotensor $S$-wave.
In Fig.~11 the vector $OC$ $({\cal F}^0 + {\cal F}^2/\sqrt{2})$ is $\sqrt{3/2}$ times the charged channel $S$-wave, $F^c$, while the vector $ON$
$({\cal F}^0 - \sqrt{2} {\cal F}^2)$ is $-\sqrt{3}$ times the neutral one, $F^n$. One sees that the square of the neutral channel $S$-wave (which dominates its cross-section) is a factor of 12 smaller than the modulus squared of the pure $I=0$ $S$-wave. It is in this (bigger) amplitude that the $\sigma$ is to be found. This delivers an $I=0$ cross-section averaged across the $\sigma$ consistent with a 4 keV width determined from the pole residue.

There are, of course, other calculations  of the two photon width of the $\sigma$ in the literature~\cite{oller}. A particularly recent one is that of  Fil'kov and Kashevarov~\cite{filkov1}. These authors fit the data in Fig.~6 with amplitudes that satisfy dispersion
relations, with a resonant shape for the imaginary part of the $\sigma$-component. In Ref.~72 they have  three fits, which determine  Breit-Wigner parameters for the $\sigma$ spanning a whole  range of values from 449 to 616~MeV  for the mass and
542 to 1655~MeV for the total width. These give corresponding  two photon widths from 0.26 to 0.99~keV, with 0.62~keV favoured in their later Ref.~71 --- much smaller than our result of Eq.~(12).
However, the treatment of Fil'kov and Kashevarov does not respect the key unitarity relation of Eq.~(3).
As just discussed in connection with Figs.~7,~10,~11, ignoring this leads to  erroneous conclusions.
 The way the phase of the $I=J=0$ component changes in the complex energy plane, Fig.~3, is crucial in the linking the pole parameters to the behaviour of the relevant amplitudes on the real energy axis from $E=0$ to $E=0.6$ GeV, in particular. 
\begin{table}[t]
\begin{center}
\label{table:1}
\renewcommand{\arraystretch}{1.15} 
\vspace{2mm}
\begin{tabular}{|c||c|c|}
\hline
\multicolumn{3}{|c|}{\rule[-0.25cm]{0cm}{7mm}{\large  $\Gamma(\gamma\gamma)$} keV }\\
\hline
&&\\[-3.mm]
composition  & predictions & author(s)\\[2.mm]
\hline
\hline
&&\\[-3.mm]
{\large (\uu + \dd)/$\sqrt{2}$} &4.0 & Babcock \& Rosner~\cite{babcock} \\[2.mm]
\hline
&&\\[-3.mm]
 {\large\st} & 0.2  & Barnes~\cite{barnes} \\[2.mm]
\hline
&&\\[-3.mm]
{\large ${\overline{[ns]}}[ns]$}, $\,n=(u,d)$ & 0.27 & Achasov {\it et al.}~\cite{achasov4q}\\ [2.mm]
\hline
&&\\[-3.mm]
& 0.6  & Barnes~\cite{barnesKK} \\[-3.mm]
{\large $\KK$} &&\\[-3.mm]
 & 0.22 & Hanhart {\it et al.}~\cite{hanhart}\\[2.mm]
\hline
\end{tabular}
\end{center}
\vspace{1.mm}
{~~~~~~~~~Table 1: Radiative widths of scalars in different modellings of their composition.}
\end{table}

What does our radiative width of $(4.1 \pm 0.3)$ keV mean for the composition of the $\sigma$? Simple predictions\cite{babcock,barnes,achasov4q,barnesKK,hanhart} are shown in Table 1. We see that the
result in Eq.~(12) seems to affirm a simple \uu, \dd composition.
However, the calculation of the $\qq$ composition is not really robust.
It can be checked by a relation provided by the naive quark model
for states in the same $L$-band. Here, 
tensors and scalars of the same quark constitution, with the same spin ($S=1$) and the same orbital angular momentum ($L=1$), satisfy~\cite{chanowitz1,chanowitz2}:
\be
\frac{\Gamma(0^{++}\to\gamma\gamma)}{\Gamma(2^{++}\to\gamma\gamma)}\;=\;\frac{15}{4}\,\left(\frac{m_0}{m_2}\right)^n \quad ,
\ee
where the exponent $n$ in the mass term depends on the relevant binding shape of the interquark potential: $n=3$ for the short-distance Coulombic component, and $n\to 0$ for the linearly confining part. With the $\sigma$ and $f_2$ masses differing by more than a factor of 2, the result is very senstive to this exponent $n$. Moreover, Li {\it et al.}~\cite{rel-corr} have estimated that this relation can be changed by up to a factor of 2 from relativistic corrections.

Alternatively, Narison~\cite{narison}
has proposed that the $\sigma$ is largely a glueball. Perturbative arguments would suggest a small two photon coupling for such a state. This is certainly difficult to reconcile with the calculated rate
of 4~keV. Once again we send out a plea for a genuine strong coupling calculation of the scalar radiative widths with different compositions
$\qq$, ${\overline{qq}}qq$ or glueball to compare with the result of Eq.~(12)
and the coupling in Eq.~(11), from which it is derived.

An important check of the calculation presented here and displayed in Figs.~7,~10
would be a more precise measurement of the charged and neutral pion differential cross-sections at low energy. This is the one part of the proposed DA$\Phi$NE physics
programme that has not yet been implemented. While the luminosity is potentially higher than either PETRA or PEP, where the Mark~II and Crystal Ball data were taken, the configuration of the DA$\Phi$NE machine requires the on-going electrons to be tagged~\cite{daphnegg1,daphnegg2,daphnegg3}. This is a feature that should be taken up in any upgraded 
DA$\Phi$NE-II. At higher energies, {\it i.e} above 800 MeV, data are now becoming available from Belle, with the charged cross-section already published~\cite{belle} and neutral data in the pipeline. These will provide results, both integrated and differential, with high statistical precision that will impact on the
first accurate determination of the $\gamma\gamma$ coupling of the higher mass scalars, the $f_0(980)$, $f_0(1370)$ and $f_0(1510)$. Detailed amplitude analyses
will be required to separate out the $I=J=0$ component of these
cross-sections cleanly, as discussed recently in Ref.~86.
In addition Belle will have data on the $\gamma\gamma\to\pi^0\eta$ channel,
which will be key to unravelling the composition of the $a_0(980)$ and $a_0(1430)$ and so learning about which of the isoscalars is their companion. Which of these, beyond the $\sigma$, make up the Higgs sector of QCD is not yet clear.

Modelled on the way the \uu, \dd condensates break chiral symmetry, 
heavy flavour condensates may similarly  break electro-weak symmetry --- see for instances Refs. 87-89.
 The first priority is to determine the location of the corresponding scalar(s). After that the correlation of couplings to the many decay channels must be found, Two photon couplings so essential  for the study of the Higgs sector of chromodynamics, are likely to prove just as important for illuminating the Higgs  of flavourdynamics  at future colliders. The radiation of photons is the key to solving the riddle of the scalars.
\vspace{3mm}

\noindent{\bf Acknowledgements}

This work was partially supported by the EU
Contracts  HPRN-CT-2002-00311, \lq\lq EURIDICE'' and 
 MRTN-2006-035482, \lq\lq FLAVIAnet''.

\end{document}